\documentclass[12pt,a4paper]{article}
\usepackage{amsmath}
\usepackage{bm}
\usepackage{amsfonts}
\usepackage{amssymb}
\usepackage{authblk}
\usepackage[round]{natbib}
\usepackage[final]{graphicx}

\usepackage[linesnumbered,ruled,vlined]{algorithm2e}
\SetKwInput{KwInput}{Input}
\SetKwInput{KwOutput}{Output}
\SetKwIF{If}{ElseIf}{Else}{if}{then}{else if}{else}{endif}

\date{}
\usepackage{setspace}
\begin{document}

\title{Big Data and model-based survey sampling}
\author[1]{Deldossi Laura}
\author[2]{Tommasi Chiara}
\affil[1]{Department of Statistical Science, Universit\`a Cattolica del Sacro Cuore, Milano}
\affil[2]{Department of Economics, Management and Quantitative Methods, University of Milan}

\maketitle

\begin{abstract} 
Big Data are  huge amounts of digital information that are automatically accrued or merged from several sources and rarely result from properly planned surveys. A Big Dataset is herein conceived of as a collection of information concerning a finite population. We suggest selecting a sample of observations to get the inferential goal.  We assume a super-population model has generated the Big Dataset. With this assumption, we can apply the theory of optimal design to draw a sample from the Big Dataset that contains the majority of the information about the unknown parameters.
\end{abstract}


\section{Introduction}
Big Data are a huge quantity of data that are automatically accrued and/or obtained by merging several sources of information.  Their availability is a great challenge currently. However, they usually arise from observational studies and not via controlled experimentation. Thus, the quality of the Big Data information might not be very good. 
In addition, in large-sample studies, if the inferential goal is to test the effect of an explanatory variable, then the p-value often leads to the rejection of the null hypothesis. That is, even very small effects can become statistically significant because of the increased power due to the huge amount of data. From here, the idea of selecting a subsample of the Big Dataset to achieve the inferential goal.
This topic has  been already studied by \cite{Ma:2015}, \citet{drovandi:2017}, \citet{Wang:2018a}, \citet{Wang:2018b} and \citet{Broderick:2019}, among others.

To accomplish this goal, first, the Big Dataset is conceived of as a finite population, even though it does not come from a population survey, since it is not a planned observation of objects (Big Data are typically generated as secondary outcomes of existing systems). This idea is not completely new and has been addressed by the research of \cite{Ma:2015} and the references therein; however, unlike the other authors, we consider analytic uses of sample surveys instead of descriptive (enumerative) uses. Analytic uses  concern inference about parameters of a specified model, which is called the super-population model, and that model is assumed to have generated the finite population values, i.e., the Big Dataset. Many books on survey sampling address the enumerative uses of surveys (such as \citet{Cochran:1977}), which are related to the estimation of summary population measures, such as the means, totals, and proportions. In descriptive surveys, the finite population parameters are of interest and inferences could be made with certainty using a census. In contrast, in a model-based sampling approach (analytic surveys), the inferences are related to the model parameters, and these parameters remain unknown, even when using a census. In this work, we make inferences about the parameters of the model that generates the Big Data, and thus we consider the model-based survey approach. In \citet{Ma:2015}, the aim instead is to use the subsample to estimate the estimator based on the full data set (which is a finite population parameter); therefore, they follow an enumerative approach. 

To form a subsample of data, we apply the theory of optimal design instead of considering the most commonly used sampling schemes. In this sense, our proposal is an answer to \cite{drovandi:2017} and their call to the experimental design community to work together in the field of Big Data analysis. In their paper, in fact, they underline \lq\lq the potential for modern decision theoretic optimal experimental design methods, which by their very nature have traditionally been applied prospectively, to improve the analysis of Big Data through retrospective designed sampling in order to answer particular questions of interest''.

Indeed, the connection between the sampling and experimental design had been already explored by \cite{Wynn:1977a}, \cite{Wynn:1977b}, \cite{Wynn:1982}, \cite{Fedorov:1989} and \cite{Pronzato:2006}, among others. These authors applied methods based on the theory of optimum experimental design to finite population sampling. They propose an equivalence theorem to identify the \lq\lq optimum'' sample $s$ of $n$ units from a population $U$ of $N$ units based on a purposive selection. They reach this goal replacing the population $U$ and the sample $s$ with probability measures $\xi_0$  and $\xi$, respectively, which are defined  on the space of the auxiliary variables $\cal X$, with $\xi$  dominated by  $\xi_0$.

We propose a purposive selection strategy – which is called the \lq\lq Optimal Design Based'' (ODB) method - consisting of two steps. First, we identify the \lq\lq most informative'' values of the explanatory variables according to an optimality criterion (these optimal \lq\lq theoretical''  values are not necessarily present in the observed Big Dataset). Then, we select the observations from the full data set that are closer to these \lq\lq theoretical'' optimal values. Hence, this \lq\lq optimal-sampling'' approach enables us to select the most \lq\lq informative'' observations from the Big Dataset. It should be noted that the second step of the ODB method is similar to step 4 of Algorithm 1 in \cite{drovandi:2017}. However, these authors proposed a Bayesian adaptive procedure. At each iteration, their method searches for the optimal experimental point by exploring the whole design space and then selects the unit of the Big Dataset closest to it.

A selection strategy that is based on D-optimality and linear models is the Information-Based Optimal Subdata Selection (IBOSS) method that was proposed by \citet{Wang:2018a}. The ODB method, unlike IBOSS, can be based on any optimality criterion (herein, we consider the D- and A-criteria) and can be applied also to non-linear models. 

Finally, we borrow the concept of \lq\lq design efficiency'' from the Optimal Design Theory as a tool to measure the quality of the Big Dataset and of the selected samples in terms of their per-unit information.  

The remainder of this paper is organized as follows. In Section 2, we recall the model-based survey sampling approach, including the notations and inferential aspects. Section 3 provides the basic definitions of optimal design theory. In Section 4, we propose our optimal design-based selection method. In Section 5, we compare the ODB approach with the IBOSS method and two classical sampling schemes: the simple random sample without replacement (SRS) and the probability proportional to size (PPS) sample. Some explanatory examples are described in Section 6 while Section 7 is concerned with our simulation study. The paper ends with a discussion and some hints for future research.

\section{Framework: model-based survey sampling}
\label{sec:2}

Before applying sampling theory to the context of Big Data, we review the standard terminology   of the model-based survey sampling approach.

A finite population $U$ is a set of $N$ units, i.e., $U=\{1,\ldots,N\}$, where $\{y_1,\ldots,y_N\}$ is the population values of a variable of interest $Y$. Herein, we are not interested in estimating the finite population parameters that are specific functions of $y_1,\ldots,y_N$ (enumerative uses of sampling). Instead, we are interested in estimating the unknown parameters of a super-population model (an approximation of the true unknown data generating process), which has generated the values $y_1,\ldots,y_N$. Let  $\mathbf{y}_U=(y_1,\ldots,y_N)$ be the realization of an $N\times 1$  random vector $\mathbf{Y}_U$ whose probabilistic law  $g(\mathbf{y}_U;\phi)$ depends on a parameter vector $\phi$. $g(\mathbf{y}_U;\phi)$ is the super-population model and a typical inferential goal could be the precise estimation of $\phi$.

The sampling mechanism is also expressed probabilistically. A sample $s$ of size $n\leq N$ is a subset of $U$. It is convenient to represent the sample $s$  using an $N\times 1$ vector $\mathbf{i}_U=(i_1,\ldots,i_N)$, where 
\begin{equation}
i_l=\left\{
\begin{array}{ll}
 1 & \mbox{ if } l\in s\\
 0 & \mbox{otherwise}
\end{array}
\right.
\qquad
 l=1,\ldots,N
\label{inclusion_ind}
\end{equation}
is the sample inclusion indicator. In other terms, the sample can be indifferently represented by $s$ or $i_U$. Let us denote $g(\mathbf{i}_U)$ as the distribution of the sample units $\mathbf{I}_U$, i.e., the probability of obtaining each of the $2^N$ possible samples from the population $U$ (the sample design). 

To use the sample data to make inferences about $\phi$, it is necessary to represent $(\mathbf{y}_U,\mathbf{i}_U)$ as the joint outcome of a random matrix $(\mathbf{Y}_U,\mathbf{I}_U)$. The joint distribution of $(\mathbf{Y}_U,\mathbf{I}_U)$ is factorized as $g(\mathbf{i}_U|\mathbf{y}_U)g(\mathbf{y}_U;\phi)$, where  $g(\mathbf{i}_U|\mathbf{y}_U)$ is the conditional distribution of $\mathbf{I}_U$, which expresses the sampling mechanism when it depends directly on the realized value $\mathbf{y}_U$ of $\mathbf{Y}_U$. Sampling schemes directly depending on the variable of interest $\mathbf{y}_U$ are called informative sampling schemes. Conversely, sampling schemes for which the joint distribution of $(\mathbf{Y}_U,\mathbf{I}_U)$ can be expressed as  $g(\mathbf{i}_U)g(\mathbf{y}_U;\phi)$ are called non-informative. 

Informative sampling schemes are rare. Schemes where the sampling depends on some other characteristics, $X_1,\ldots,X_p$, of the population are instead quite common. Let $\mathbf{X}=(X_1,\ldots,X_p)$ be the $p\times 1$ vector of the auxiliary characteristics (or design variables), and we will refer to $\mathbf{X}$ as the experimental condition.  $\mathbf{x}_U$ denotes the $N\times p$ population design matrix whose rows are $\mathbf{x}_1,\ldots,\mathbf{x}_N$, i.e., the population values of $\mathbf{X}$. The population auxiliary matrix $\mathbf{x}_U$ can be viewed as a realization of a random matrix $\mathbf{X}_U$ with the distribution $g(\mathbf{x}_U;\psi)$. To stress the dependence of the sample design on $\mathbf{x}_U$, the sampling mechanism is expressed as a conditional distribution $g(\mathbf{i}_U|\mathbf{X}_U=\mathbf{x}_U)$. If there is no additional direct dependence of sampling on $\mathbf{y}_U$, then the joint distribution of $(\mathbf{Y}_U,\mathbf{I}_U)$ conditional on $\mathbf{X}_U=\mathbf{x}_U$ can be expressed as $g(\mathbf{i}_U|\mathbf{X}_U=\mathbf{x}_U)g(\mathbf{y}_U|\mathbf{X}_U=\mathbf{x}_U;\theta)$. This assumes that $\mathbf{Y}_U$ and $\mathbf{I}_U$ are conditionally independent given $\mathbf{X}_U=\mathbf{x}_U$. Then, the sampling scheme is said to be non-informative conditional on $\mathbf{x}_U$. In this case, the conditional distribution  $g(\mathbf{y}_U|\mathbf{X}_U=\mathbf{x}_U;\theta)$ plays the role of the super-population model and the inferential aim is the estimation of $\mathbf{\theta}\in \Theta\subseteq {\rm I\!R}^m$.

Let us again stress the main difference between design-based and model-based approaches to survey sampling. Under design-based inferences, the only source of random variation is that which is induced in the vector $\mathbf{i}_U$ by the sampling mechanism, which is a known probability sample design. The vector of finite population values $\mathbf{y}_U$ is treated as fixed, thereby avoiding the need to specify a model that generates it, and the descriptive inference is the traditional setting. Models may be used in this approach just to motivate the choice of estimators, and, in this case, the approach is called model-assisted.

In the model-based approach, if the sampling scheme is non-informative, the model that generates $\mathbf{y}_U$ is the only source of random variation to be taken into account, and inferences may be obtained using the maximum likelihood. In more detail, let us suppose that the rows of $\mathbf{y}_U$ corresponding to the sampled units form the $n\times 1$ vector $\mathbf{y}_{s}$. Supposing that $\mathbf{i}_U$, $\mathbf{y}_{s}$ and $\mathbf{x}_U$ are observed while  $\mathbf{y}_{\bar s}$ is unobserved, the data consist of $(\mathbf{i}_U,\mathbf{y}_{s},\mathbf{x}_U)$ and the likelihood for $(\theta,\psi)$ is given by
$$
{\cal L}(\theta,\psi) \propto \int g(\mathbf{i}_U|\mathbf{y}_U,\mathbf{x}_U)g(\mathbf{y}_U|\mathbf{x}_U;\theta)g(\mathbf{x}_U;\psi)\,d\mathbf{y}_{\bar s}.
$$
If sampling is non-informative given $\mathbf{x}_U$, then the term $g(\mathbf{i}_U|\mathbf{y}_U,\mathbf{x}_U)=g(\mathbf{i}_U|\mathbf{x}_U)$ and it can be dropped from the likelihood since it is a constant with respect to $(\theta,\psi)$, i.e.
\begin{equation}
{\cal L}(\theta,\psi)\propto  g(\mathbf{x}_U;\psi) \int g(\mathbf{y}_U|\mathbf{x}_U;\phi)\,d\mathbf{y}_{\bar s}.
\label{likelihood}
\end{equation}
This means that the sampling can be ignored for the likelihood-based inference about $(\theta,\psi)$; in other words, we may treat sample $s$ or $\mathbf{i}_U$ as fixed (see \citet{Chambers:2013}, p. 8). If the parameter of interest is only $\theta$, then, from (\ref{likelihood}), the corresponding pseudo-likelihood (it is a conditional likelihood) is  
\begin{equation}
\label{likelihood_2}
{\cal L}(\theta) \propto  \int g(\mathbf{y}_U|\mathbf{x}_U;\theta)\,d\mathbf{y}_{\bar s} = g(\mathbf{y}_s|\mathbf{x}_U;\theta).
\end{equation}

Hence, following \cite{Chambers:2012} (page 10), a sampling method is non-informative for performing inferences about a parameter of the super-population model if the model that is specified at the population level also applies to all sampled observations, regardless of what is selected.

\subsection{Big Data as a finite population}
A Big Dataset is an $N\times (p+1)$ matrix that contains $N$  observations of $p+1$ variables. Usually, $N$ is very high, while $p$ may be high or not, depending on the context from which the Big Data originate. Since the information that is recorded in a Big Dataset concerns $N$ statistical units, it can be considered as a “finite population” that is generated from a super-population model.

Let us assume that the first column of the Big Dataset concerns the information on a variable of interest $Y$ and the remaining columns contain information on some other auxiliary variables $X_1,\ldots,X_p$. In short, the Big Dataset can be written as $[\mathbf{y}_U,\mathbf{x}_U]$. Let us assume hereafter that we have already removed the less informative auxiliary variables from the Big Data using, for instance, the method that was proposed by \cite{Fedorov:2006}, in addition to other techniques. 

Expression (\ref{likelihood}) justifies our idea of using a subset of the Big Data to make inferences about $\theta$ because the functional form of the likelihood is the same for the whole Big Dataset $g(\mathbf{y}_U|\mathbf{x}_U;\theta)$ and for the sample data $g(\mathbf{y}_s|\mathbf{x}_U;\theta)$. The only difference is in the number of observations, which is $N$ in the first case and $n$ in the latter. 
As a consequence,  for any non-informative sampling design, if we use the maximum likelihood estimator 
of $\theta$, then the sampling mechanism can be ignored when making inferences.

 However, when studying the properties of other kind of estimators (see, for instance, \citet{Ma:2015}), 
 it is necessary to consider both sources of variation: the sampling design and the super-population model.


Let us note that the observed values of the design variables, which form $\mathbf{x}_U$, explain any difference between the sampled and non-sampled units through $g(\mathbf{i}_U|\mathbf{x}_U)$. Hence, the population auxiliary matrix $\mathbf{x}_U$ may be used to assist in the selection of the sample. This suggests that there is a connection between the optimal design of experiments and model-based survey sampling. 

In the next section, we briefly review the main concepts and definitions in the theory of optimal experimental designs.

\section {Overview on optimal experimental design}
\label{sec:3}

Experimental design is a guided process of collecting data as effectively as possible to reach an inferential goal (precise estimation of some parameters, model discrimination, prediction or any other objective).
 
Let $\mathbf{x}\in {\cal X}$ be an experimental condition that can be chosen by the experimenter in an experimental domain ${\cal X}\subseteq {\rm I\!R}^p$ and $y=y(\mathbf{x})$ be the corresponding response variable.

Let $\mathbf{y}_U=(y_1,\ldots,y_N)$ be a vector of $N$ independent responses that are observed at $N$ and not necessarily different experimental conditions $\{\mathbf{x}_1, \ldots, \mathbf{x}_N\}$. This set of experimental points is called the exact design. When some experimental conditions are repeated, then the exact design can be defined using the distinct experimental points $\mathbf{x}_1, \ldots, \mathbf{x}_k$ and the number of observations $n_j$ to be taken at each $\mathbf{x}_j$, $j=1,\ldots,k<N$.

Let us assume that the responses and experimental conditions are related (at least approximatively) through the following linear regression model:
$$
\mathbf{y}_U \cong \mathbf{F}_U \,\theta+\varepsilon_U, 
$$
where $\theta=(\theta_0,\ldots,\theta_m)^T$ is an $(m+1)\times 1$ vector of unknown parameters of interest, $\varepsilon_U=(\varepsilon_1,\ldots,\varepsilon_N)^T$ is a vector of homoskedastic independent errors such that ${\rm E}(\varepsilon_i)=0$ and ${\rm Var}(\varepsilon_i)=\sigma^2$, and
$$
\mathbf{F}_U= \left[\!\!
   \begin{array}{c}
    \mathbf{f}(\mathbf{x}_1)^T   \\
    \vdots\\
    \mathbf{f}(\mathbf{x}_N)^T
   \end{array}
   \!\!\right]\quad \mbox{with} \quad \mathbf{f}(\mathbf{x})^T=(f_1(\mathbf{x}),\ldots,f_m(\mathbf{x}))
   $$
is the design matrix and $f_1(\mathbf{x}),\ldots,f_m(\mathbf{x})$ are $m$ linear independent known functions (see for instance, \citet{Pazman:1986}). It is well known that the precision matrix of the BLUE for $\theta$ is
\begin{equation}
\frac{1}{\sigma^2} \mathbf{F}_U^T \mathbf{F}_U =\frac{1}{\sigma^2}\sum_{i=1}^N \mathbf{f}(\mathbf{x}_i) \mathbf{f}(\mathbf{x}_i)^T
  = \frac{N}{\sigma^2} M(\xi_N),
\end{equation}
where
$$
 M(\xi_N)= \sum_{j=1}^k \mathbf{f}(\mathbf{x}_j) \mathbf{f}(\mathbf{x}_j)^T \frac{n_j}{N}
 =\frac{\mathbf{F}_U^T \mathbf{F}_U}{N}
$$
is called the information matrix of the exact design $\xi_N$ and it plays a crucial role in the theory of optimal experimental design. Our aim is to collect the \lq\lq optimal'' data in order to provide a precise estimation of $\theta$, which means to find an exact design $\xi_N^*$ that maximizes some concave function $\Phi[M(\xi_N)]$.  Computing the exact optimum designs can be difficult because they are solutions of finite programming problems. For this reason, the definition of a continuous design was introduced as a discrete probability measure on ${\cal X}$ with a finite number of support points:
$$\xi=\left\{\!\!
\begin{array}{ccc}  
\mathbf{x}_1 & \cdots & \mathbf{x}_k \cr  
\omega_1     & \cdots & \omega_k  
\end{array}\!\!
\right\},\;\; 0\leq \omega_j\leq 1,  \;\sum_{j=1}^{k}\!\omega_j=1. 
$$
The information matrix for a continuous design $\xi$ is
$
M(\xi)=\sum_{j=1}^k \mathbf{f}(\mathbf{x}_j) \mathbf{f}(\mathbf{x}_j)^T\: \omega_j.
$
A continuous optimum design, 
\begin{eqnarray*}
\xi^*=\arg\max_{\xi} \Phi[M(\xi)],
\end{eqnarray*}
can be found as a solution to a differential problem. Even when a continuous optimum design cannot be applied in practice, such as when $N\omega_j$ is not an integer, $\xi^*$ can be used as a benchmark for any other design. 

The functional $\Phi[\cdot]$ summarizes the inferential goal and is called the optimality criterion. In this paper, we consider the well-known D- and A-optimality criteria, i.e., $\Phi_D[M(\xi)]=|M(\xi)|$ and $\Phi_A[M(\xi)]=-{\rm Tr}\, \big[M(\xi)^{-1}\big]$, respectively. 
The D-optimum design minimizes the generalized variance of $\theta$ as
$\xi_{D}^*=\arg\max_{\xi}\Phi_D[M(\xi)]=\arg\min_{\xi} |M(\xi)^{-1}|$. The A-optimum design minimizes the total variation of $\theta$ as $\xi_{A}^*=\arg\max_{\xi}\Phi_A[M(\xi)]=\arg\min_{\xi} {\rm Tr}\,\big[M(\xi)^{-1}\big]$. 
Since the generalized variance and total variation are common overall dispersion measures of a random vector, the D- and A-optimality criteria are frequently applied. Other choices of $\Phi[\cdot]$, however, are possible, that reflect different inferential purposes.

If $\phi[\cdot]$ is an homogeneous function, i.e., $\Phi[\bm{M}(c\,\xi)]=\Phi[\bm{M}]/c$ where $c$ is a constant (see \cite{Pazman:1986}), then a measure of the goodness of $\xi$ with respect to  $\xi^*$ is
\begin{equation}
 0\leq {\rm Eff}_{\Phi}[\bm{M}(\xi)]=\frac{\Phi[\bm{M}(\xi)]}{\Phi[\bm{M}(\xi^*)]}\leq 1. 
 \label{efficiency}
\end{equation}
which is called the design efficiency of $\xi$. If an optimality criterion is not homogeneous, then the corresponding efficiency is defined using an equivalent homogeneous criterion. We recall that for D- and A-criteria the efficiencies are ${\rm Eff}_{D}(\xi)=\left(|\bm{M}(\xi)|/|\bm{M}(\xi^*)|\right)^{1/m}$ and ${\rm Eff}_{A}(\xi)={\rm Tr}\big[\bm{M}(\xi^*)^{-1}\big]/{\rm Tr}\big[\bm{M}(\xi)^{-1}\big]$, respectively.

\section{A sampling rule based on  an optimality criterion}
\label{sec:4}

We assume that the Big Data $[\mathbf{y}_U,\mathbf{x}_U]$ are generated by a super-population model as follows:
\begin{equation}
y_i   \cong  \mathbf{f}(\mathbf{x}_i)^T\theta+\varepsilon_i,\quad
{\rm E}(\varepsilon_i)=0,\; {\rm Var}(\varepsilon_i)=\sigma^2, \quad i=1,\ldots,N.
\label{superpop_model}
\end{equation}
We are interested in estimating the parameter vector $\theta$ as precisely as possible. 

Big Data usually come from an unplanned observation of values (numerical or not) that are not controlled by the observer. The matrix $\mathbf{x}_U$ may be considered as the observed experimental design. We can consider this as \lq\lq Nature'' applying the treatments $\mathbf{x}_U$ and the researcher can only observe the consequences $\mathbf{y}_U$. If Nature had been a \lq\lq wise'' experimenter, then it would have chosen the $N$ values for explanatory variable $\mathbf{X}$ according to an optimality criterion $\Phi[\cdot]$. In other words, given a super-population model, it is always possible to compute the (continuous) optimum design as
\begin{equation}
\xi^*=\arg\max_{\xi} \Phi[M(\xi)]=\left\{\!\!
\begin{array}{ccccc}  
\mathbf{x_1^*}& \cdots&  \mathbf{x_j^*}& \cdots & \mathbf{x_k^*}        \cr  
\omega_1^*    & \cdots & \omega_j^*    & \cdots & \omega_k^*  
\end{array}\!\!
\right\}. 
\label{phi-optimum}
\end{equation}
A wise Nature would have generated $N \omega_j^*$ responses at $\mathbf{x_j^*}$, $j=1,\ldots,k$ and this would have been the \lq\lq ideal''  combination of treatments $\mathbf{x}_U^*$ that Nature should have applied to obtain a precise parameter estimation.

Let $\bf{F}_U=\bf{F}_U(\mathbf{x}_U)$ be the observed design matrix and  $\bf{F}_U^*=\bf{F}_U(\mathbf{x}_U^*)$ the ideal design matrix. $\bf{M}(\mathbf{x}_U)=\bf{F}_U^T\bf{F}_U/N$ measures the per-unit information contained in the Big Data and  from (\ref{efficiency}) the quality of the Big Data can be measured by

 \begin{equation} 
 0\leq {\rm Eff}_{\Phi}[\bm{M}(\bm{x}_U)]=\frac{\Phi[\bm{F}_U^T \bm{F}_U]}{\Phi[{\bm{F}_U^*}^T \bm{F}_U^*]}\leq 1.
 \label{Eff}
 \end{equation}

Low efficiency  means that the Big Data  contains poor information for precisely estimating the parameters of the super-population model.

It is well known that standard statistical methods for parameter estimations are not applicable with very huge data sets because of computational limitations.  
Let $\bm{F}_s$ denote the $n\times (m+1)$ design matrix corresponding to a sample $s$ of $n$ units, and $\bm{F}_s$ is obtained from $\bm{F}_U$ by removing the rows corresponding to  $\bar s$ (set of not sampled units). In addition, let $\bm{M}_s=\bm{F}_s^T \bm{F}_s/n$ be the sample information matrix that measures the per-unit information that is contained in $s$. The \lq\lq goodness'' of a sample $s$ can be still measured by the efficiency (\ref{efficiency}): 
\begin{equation}
   {\rm Eff}_\Phi[\bm{M_s}]=\frac{n\Phi[\bm{F}_s^T \bm{F}_s]}{N\Phi[{\bm{F}_U^*}^T \bm{F}_U^*]}. 
   \label{Efficiency}
\end{equation}
The most \lq\lq informative'' sample, say $s^*$, provides the largest per-unit sample information, i.e.,  $\Phi[\bm{F}_{s^*}^T \bm{F}_{s^*}]=\arg\max_s \Phi[\bm{F}_{s}^T \bm{F}_{s}] $. 
Unfortunately $s^*$ cannot be computed because $N$ is a huge value. It is not  possible to select all the 
${N \choose n}$ samples and  compute all the corresponding $\Phi[\bm{F}_s^T \bm{F}_s ]$. 

Approximations of $s^*$ can be found by applying, for instance, the exchange algorithm (two references are \citet{Mitchell:1970} and \citet{Wynn:1972}, among others), but it is computationally expensive. We instead propose a sampling rule that is based on the continuous $\Phi$-optimum design that can be computed quite easily despite the magnitude of $N$. More specifically, since the $\Phi$-optimum design (\ref{phi-optimum}) gives the best combination of treatments for the precise estimation of $\theta$, we suggest selecting the $n \omega_j^*$ rows of $\mathbf{F_U}$ that are the closest to  $\mathbf{f(\mathbf{x_j^*})}^T$ for $j=1,\ldots,k$. As a measure of closeness, the Euclidean, Mahalanobis or any other distance can be applied. When $n \omega_j^*$ is not an integer number, then a suitable rounding-off rule can be applied (see for instance \citet{Pulkesheim:1992}). We call this sampling rule the optimal design based (ODB) sampling scheme. This is shown in Algorithm 1.
\begin{algorithm}{
\caption{Selection of the ODB sample }
\label{Algorithm 1}
 \SetAlgoLined
 \vspace{0.1cm}
 \KwInput{Big Dataset [$\mathbf{y_U}, \mathbf{x_U}$], sample size $n$, design matrix $\mathbf{F}$, optimality criterion $\Phi[\cdot]$} 
 \KwOutput{ODB sample}
 Through the function od.AA in \cite{OptimalDesign}, compute the design $\xi^*=\arg\max_{\xi} \Phi[M(\xi)]$ (see eq. \ref{phi-optimum}) and the corresponding ideal design matrix $\mathbf{F^*_U=F_U(x^*_U)}$\;
 \If{$n\omega_j^*$ for $j=1, \cdots,k$ is not an integer number} {
  applying the rounding multiplier rule proposed by Fedorov \citep{Pulkesheim:1992} compute $n_j=\lceil(n-k)\omega_j^*\rceil$ and  $n^*_j=n\omega_j^*-n_j$, for $j=1, \cdots,k$, where the function  $\lceil \cdot \rceil$ means to round up to the next integer\;
  compute $\tilde n=n-\sum_{j=1}^k n_j$\;
  let $\{r_1,r_2, \ldots,r_k\}$ be the ranks of $\{n^*_1, \ldots, n^*_k\}$ arranged in descending order\;
  let $n_c$ be the number of centroids $\mathbf{x_j^*}$ such that $n^*_j>0$\;
  \While{$\tilde n>0$}
  {\eIf{$\tilde n \leq n_c$}{
  $n_{r_i}=n_{r_i}+1$ for $i=1, \cdots, \tilde n$\;
  }{
  $n_{r_i}=n_{r_i}+1$ for $i=1, \cdots,n_c$ \;}
  let $\ddot n_j$ be the update value of $n_j$ for $j=1, \cdots,k$\; 
  compute $\tilde n=n-\sum_{j=1}^k \ddot n_j$\;
  }
 }
  \For {$j=1,\ldots, k$}{compute the distance $||\mathbf{f(x_j)^T}-\mathbf{f(x_j^*)^T}||$ \;
  let $\{d_1,d_2, \ldots,d_N\}$ be the ranks of $||\mathbf{f(x_j)^T}-\mathbf{f(x_j^*)^T}||$ arranged in ascending order\;
  select from $\mathbf{F_U}$ the rows $d_1, \cdots, d_{\ddot n_j}$\;
  }
  }
\end{algorithm}

\section{Comparing non-informative samples}
\label{SEC:comparing}

As stressed in Section 2, when the sampling mechanism is non-informative, then the likelihood (\ref{likelihood_2}) can be applied to make inferences about the unknown parameter vector. Let 
\begin{eqnarray*}
    \hat \theta_s 
    & = & (\mathbf{F}_s^T \mathbf{F}_s)^{-1}\mathbf{F}_s \mathbf{y}_s\\
    & = & \left(\sum_{l=1}^N \mathbf{f}(x_l)\mathbf{f}(x_l)^T i_l\right)^{-1} \sum_{l=1}^N \mathbf{f}(x_l)^T y_l\, i_l
\end{eqnarray*}
denote the ordinary least square (OLS) estimator based on the sampled units, where $i_l$ is the inclusion indicator that is defined in (\ref{inclusion_ind}). It is well known that $\hat \theta_s$ is unbiased with the covariance matrix that is given by $\sigma^2 (\mathbf{F}_s^T \mathbf{F}_s)^{-1}$. If the most informative sample $s^*$ could be identified, then given an optimality criterion $\Phi$, $\Phi(\mathbf{F}_{s^*}^T \mathbf{F}_{s^*})^{-1}\leq \Phi(\mathbf{F}_{s}^T \mathbf{F}_{s})^{-1}$. As a consequence, since we aim to approximate $\Phi(\mathbf{F}_{s^*}^T \mathbf{F}_{s^*})^{-1}$ using the ODB method, this method generally provides a more precise estimate than other samples. In the literature, to compare different sampling strategies, the marginal properties (with respect to the inclusion indicator, i.e., the random selection) of the estimators are usually considered. The OLS estimator is unbiased as follows: 
$$
{\rm E}_{\mathbf{I}_U,\mathbf{Y}}(\hat \theta_s|\mathbf{X}_U)=
{\rm E}_{\mathbf{I}_U}\left[ {\rm E}_{\mathbf{Y}} (\hat \theta_s|\mathbf{X}_U) \right]=\theta,
$$
where the second equality is due to the non-informativeness of the sampling strategy.
In addition,
\begin{eqnarray*}
{\rm Var}_{\mathbf{I}_U,\mathbf{Y}}(\hat \theta_s|\mathbf{X}_U)&=&
{\rm E}_{\mathbf{I}_U}\left[ {\rm Var}_{\mathbf{Y}} (\hat \theta_s|\mathbf{X}_U) \right]+{\rm Var}_{\mathbf{I}_U}\left[ {\rm E}_{\mathbf{Y}}(\hat \theta_s|\mathbf{X}_U)\right]
\nonumber \\
&=& \sigma^2 {\rm E}_{\mathbf{I}_U}\!\!\left[\! \left(\sum_{l=1}^N \mathbf{f}(x_l)\mathbf{f}(x_l)^T i_l\right)^{-1} \right]
=\sigma^2\, {\rm E}_{\mathbf{I}_U}\!\!\left[\!\left( \mathbf{F}_s^T \mathbf{F}_s   \right)^{-1} \right].
\end{eqnarray*}
If the optimality criterion function $\Phi$ is concave or linear (as for D- and A-criteria, respectively), then 
$$
\Phi\left\{ {\rm E}_{\mathbf{I}_U}\!\!\left[ \left( \mathbf{F}_s^T \mathbf{F}_s   \right)^{-1} \right]  \right\}
\geq 
{\rm E}_{\mathbf{I}_U} \left\{ \Phi\left[ \left( \mathbf{F}_s^T \mathbf{F}_s   \right)^{-1} \right] \right\}
\geq
\Phi\left[ \left( \mathbf{F}_{s^*}^T \mathbf{F}_{s^*}   \right)^{-1} \right]. $$
Hence, the ODB method is still expected to not be inferior to any other non-informative sampling technique. 

\section{Explanatory examples}
\label{SEC:Explanatory}

 In this section we introduce two simple examples to show how our proposal works in the selection of $n=120$ observations from a full data set of $N=10.000$ items.
We consider the super-population model (\ref{superpop_model}) with $p=2$ independent explanatory variables $X_1$ and $X_2$ and $f(\mathbf{x})^T=(1,x_1,x_2, x_1^2, x_2^2, x_1 \cdot x_2)$.

We compare the following sampling methods.
\begin{enumerate}
    \item 
     The simple random sampling in which the related information matrix is denoted by $\bm{M}_{SRS}$.
    \item
     The PPS sampling with selection probabilities that are given by  
     $$p_i=\frac{\bm{f}(\bm{x}_i)^T (\bm{F}_U^T \bm{F}_U)^{-1} \bm{f}(\bm{x}_i)}{m+1}, \quad i=1,\cdots,N$$
     to more frequently select the units with the largest prediction variance (see also \citet{Ma:2015}). The corresponding design matrix is denoted by $\bm{M}_{PPS}$.
    \item The exchange algorithm, which proceeds as follows.
    \begin{enumerate}
     \item 
     An initial sample of rows $s_0$ is chosen at random  from the $N$ rows of the Big Dataset. \item
     $s_0$ is  improved by adding that row $\bm{x}$ which  most improves the $\Phi$-criterion, followed by removing that row from $s_0 \cup \{\bm{x}\}$, which gives the  smallest reduction in the $\Phi$-criterion. \item 
     This add/remove procedure is continued until it converges, with the same row being added and then removed.
    \end{enumerate}
The corresponding information matrix is denoted by $\bm{M}_{EA}$.
    \item The IBOSS that was introduced by \citet{Wang:2018a}, where the related information matrix is $\bm{M}_{IBOSS}$.
    \item 
    The ODB method based on the D- or A-optimality criteria, where  the corresponding information matrix is denoted by $\bm{M}_{ODB}$.
\end{enumerate} 
\begin{figure}[!ht]
\vspace{6pc}
\caption{The full data set of size $N=10.000$ and the subsamples of size $n=120$ when $X_1$ and $X_2$ are independently distributed as $U(-1,1)$ and D-optimality is used in the exchange algorithm and ODB methods.}
\label{fig:1}
\includegraphics[width=0.99\textwidth]{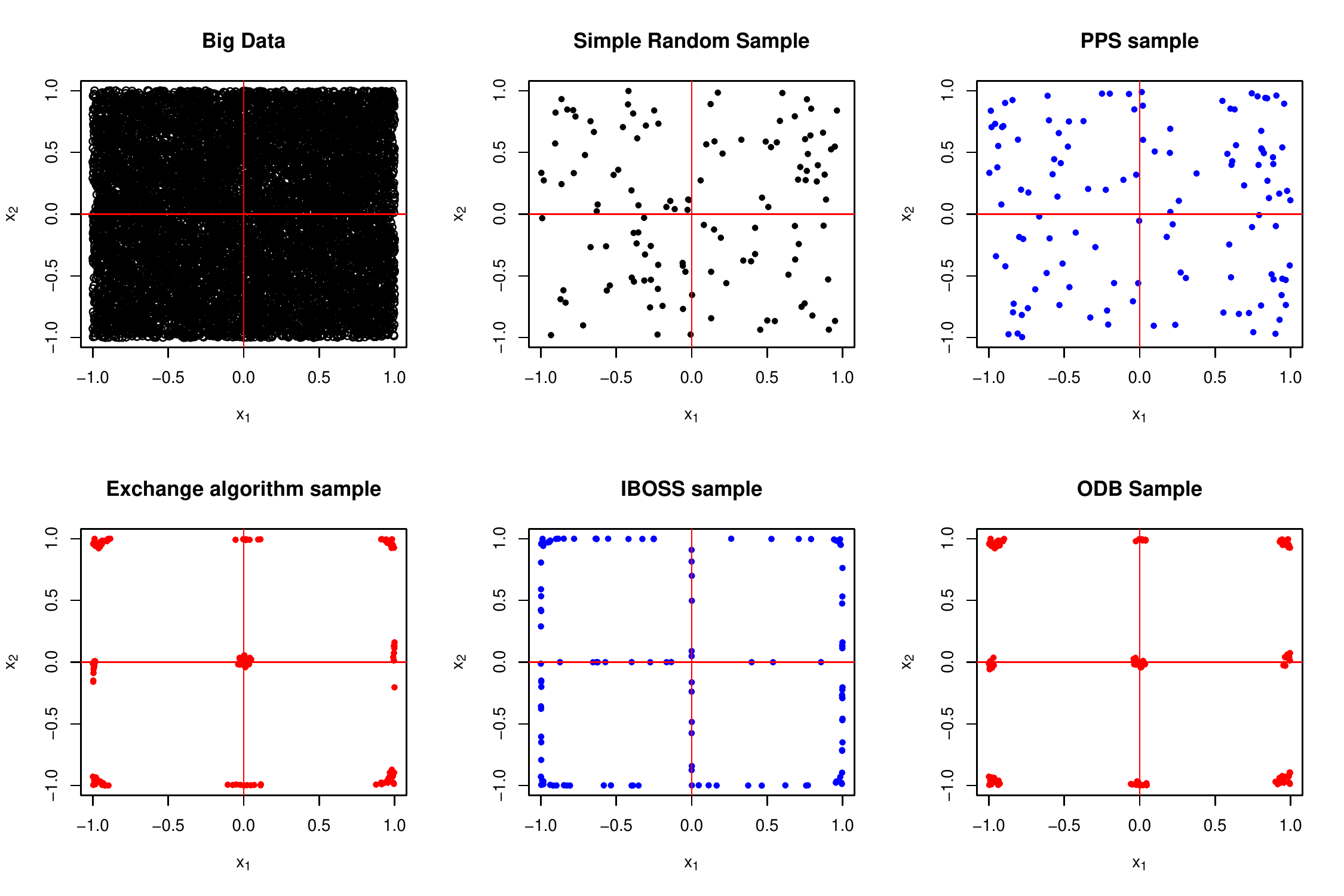}
\end{figure}
In Figure \ref{fig:1}, we assume that both $X_1$ and $X_2$ are independently distributed as $U(-1,1)$ and we specify the D-optimality in the exchange algorithm and ODB methods. The D-efficiencies at the different information matrices are as follows:  
\begin{eqnarray*}
&& {\rm Eff}_D(\bm{M})=0.4559,\quad {\rm Eff}_D(\bm{M}_{SRS})=0.4609,\quad {\rm Eff}_D(\bm{M}_{PPS})=0.5559,\\
&& {\rm Eff}_D(\bm{M}_{EA})=0.9196, \quad {\rm Eff}_D(\bm{M}_{IBOSS})=0.8331, \quad {\rm Eff}_D(\bm{M}_{ODB})=0.914.
\end{eqnarray*}
The results for the same setting and A-optimality are reported in the Supplementary materials (see Figure \ref{fig:5}).

\begin{figure}[!ht]
\vspace{6pc}
\caption{The full data set of size $N=10.000$ and the subsamples of size $n=120$ when $(X_1,X_2) \sim N(\mathbf{0},\Sigma)$ 
$ \Sigma\!=\!$
and 
A-optimality is used in the exchange algorithm and ODB methods.}
\label{fig:2}
\includegraphics[width=0.99\textwidth]{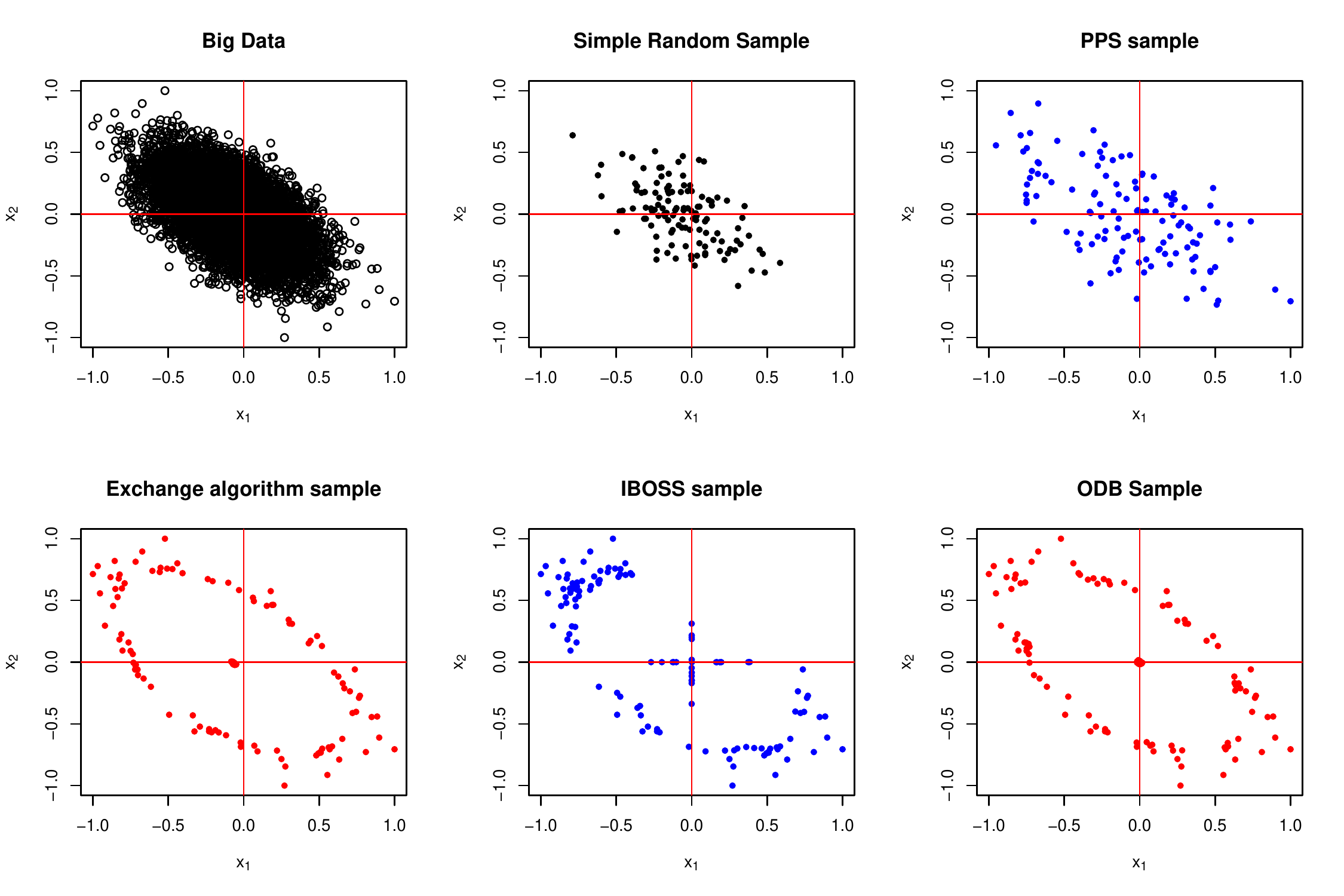}
\end{figure}

As a second example, we assume that $(X_1,X_2) \sim N({\bf{0}},\Sigma)$ with 
$ \Sigma\!=\!
  \left[\!\!\!\begin{array}{cc}
  16\!\!\! &\!\-9 \cr
  -9 \!\!\!&\!16 \cr
   \end{array}\!\!\!\right]$
and we consider the A-optimality criterion in the exchange algorithm and ODB methods (see Figure \ref{fig:2}, where the data are plotted after a suitable transformation in the square $[-1,1]\times[-1,1]$). 

The A-efficiency at the different information matrices are as follows:  
\begin{eqnarray*}
&& {\rm Eff}_A(\bm{M})=0.0121,\quad {\rm Eff}_A(\bm{M}_{SRS})=0.0112,\quad {\rm Eff}_A(\bm{M}_{PPS})=0.0384,\\
&& {\rm Eff}_A(\bm{M}_{EA})=0.0902, \quad {\rm Eff}_A(\bm{M}_{IBOSS})=0.0502, \quad {\rm Eff}_A(\bm{M}_{ODB})=0.0896.
\end{eqnarray*}

The results for the same setting and D-optimality are reported in the Supplementary Materials (see Figure \ref{fig:6}).

In both of the previous examples (see also Figures \ref{fig:7}, \ref{fig:8}, \ref{fig:9}, and \ref{fig:10} in the Supplementary Materials), the efficiencies show that SRS is the less efficient method for obtaining an informative subsample. We achieve an improvement if we apply the PPS strategy, but the best results are obtained when we consider the exchange algorithm and the IBOSS and ODB methods, which are based on the specific goal of extracting the most informative observations. 

The exchange algorithm performs better than the other two sampling schemes because it searches for the best observations - conditional to the chosen criterion function - among the available data set, regardless of the range of $(X_1,X_2)$, while the ODB method selects the observations that are near to the \lq\lq optimal points'' in the square $[-1,1]\times[-1,1]$. Unfortunately, however, the standard exchange algorithm (that we have applied) crashes when $N$ and/or $n$ increase. The development of new and more computational feasible exchange algorithms will be a matter of future research.  

Note that the efficiency is useful for both identifying the best sampling scheme and evaluating the quality of the Big Dataset. Herein, this quality is measured with respect to the “theoretical” most informative $(X_1,X_2)$-observations in the square $[-1,1]\times[-1,1]$. In the second example, the actual range is an ellipsis and not a square. For this reason, all the efficiencies are very small, which is true even when we apply the exchange algorithm and the IBOSS and ODB methods.

\section{Simulation results}
\label{sec:7}

The goal of this simulation study is to empirically compare the four different sampling strategies that were introduced in Section \ref{SEC:Explanatory}. Consistently with the theoretical results that were described in Section \ref{SEC:comparing}, we work conditional to $\mathbf{X}_U$ and thus we generate just one  $N \times (p+1)$ design matrix $\mathbf{X}_U$, where $N=10^6$, $p=10$ and $X_i$ are independently distributed as $U(0,1)$ for $i=1, \cdots,10$.\\

\vspace{0.25cm}
{\bf Simulation 1}\\
A response vector of size $N$ is simulated $R=1000$ times from the super-population linear model (\ref{superpop_model}) with  $\mathbf{f}(\mathbf{x})=(1,x_1,x_2, \cdots, x_{10})^T$,   $\bm{\theta}=(2,.5,1,1,1,2,2,2,4,4,4)^T$ and $\sigma^2=9$.

In this way, we have $R$ simulated Big Datasets. Let $r=1, 2, \cdots,R$ be the index of a simulation step.

At each step $r$, a subsample $s$ of size $n=200$ is drawn from the Big Dataset according to the four different sampling schemes.

Actually, at each simulation step, the ODB and IBOSS methods always provide the same subsample because, given $\mathbf{X}_U$, they are deterministic. SRS and PPS are different in that they are random selection methods, so we draw $100$ different SRS and PPS independent samples at each step $r$.

For each subsample, we compute the OLS estimates of the coefficients in the linear model and the D- and A-efficiencies according to equation (\ref{Efficiency}). The Monte Carlo covariance matrix of these subsample estimates provides an estimation of the covariance matrix $\sigma^2\, {\rm E}_{\mathbf{I}_U}\!\!\left[\!\left( \mathbf{F}_s^T \mathbf{F}_s   \right)^{-1} \right]$. 

Table~\ref{tab:Efficiency} gives the Monte Carlo averages of the D- and A-efficiencies of the Big Dataset and of the subsamples that are obtained from the different strategies: ODB, IBOSS, SRS and PPS, respectively.
\begin{table*}[!h]
    \caption{The Monte Carlo averages of D- and A-efficiencies of the Big Dataset and the subsamples that are obtained using ODB, IBOSS, SRS and PPS.}
\label{tab:Efficiency}

\centering
\begin{tabular}{c|cccccc}
{\bf $\Phi$-Efficiency} & Big Data &  ODB & IBOSS & SRS & PPS\\
\hline
         $D$-Efficiency    & 0.3684 &  0.6170 & 0.4246 & 0.3584 & 0.3821\\
         $A$-Efficiency    & 0.3549 &  0.5998 & 0.4025 & 0.3357 & 0.3594\\
\end{tabular}
\end{table*}

Table~\ref{tab:D_A} lists the determinants and the traces of the Monte Carlo covariance matrices of the estimates that are obtained from the entire Big Dataset and from the subsamples that are selected using the different strategies. As expected, the estimates that are obtained using the entire dataset are the most precise. Instead, if we need to select a subsample, it is preferable to avoid the SRS or PPS strategies, as shown in the last two columns of Table~\ref{tab:D_A}.
\begin{table*}[!h]
 \caption{The determinant and trace of the Monte Carlo covariance matrix of the estimates of the Big Dataset and the subsamples that are obtained using ODB, IBOSS, SRS and PPS.}
\label{tab:D_A}
\centering
    \begin{tabular}{c|cccccc}
         {\bf Criterion} & Big Data &  ODB  & IBOSS & SRS & PPS\\
         \hline
         Determinant  & 1.4e-45 & 2.4e-07 & 1.7e-05 & 1.8e-04 & 8.6e-05\\
         Trace  & 0.0013  &  3.9213 & 5.8610 & 7.2002 & 6.6926\\
    \end{tabular}
\end{table*}

\begin{figure}[!ht]
\vspace{6pc}
\caption{The boxplots of the parameter estimates that are obtained in Simulation 1 using the Big Dataset and the subsamples are selected using the ODB, IBOSS, SRS and PPS methods.}
\label{fig:3}
\includegraphics[width=0.99\textwidth]{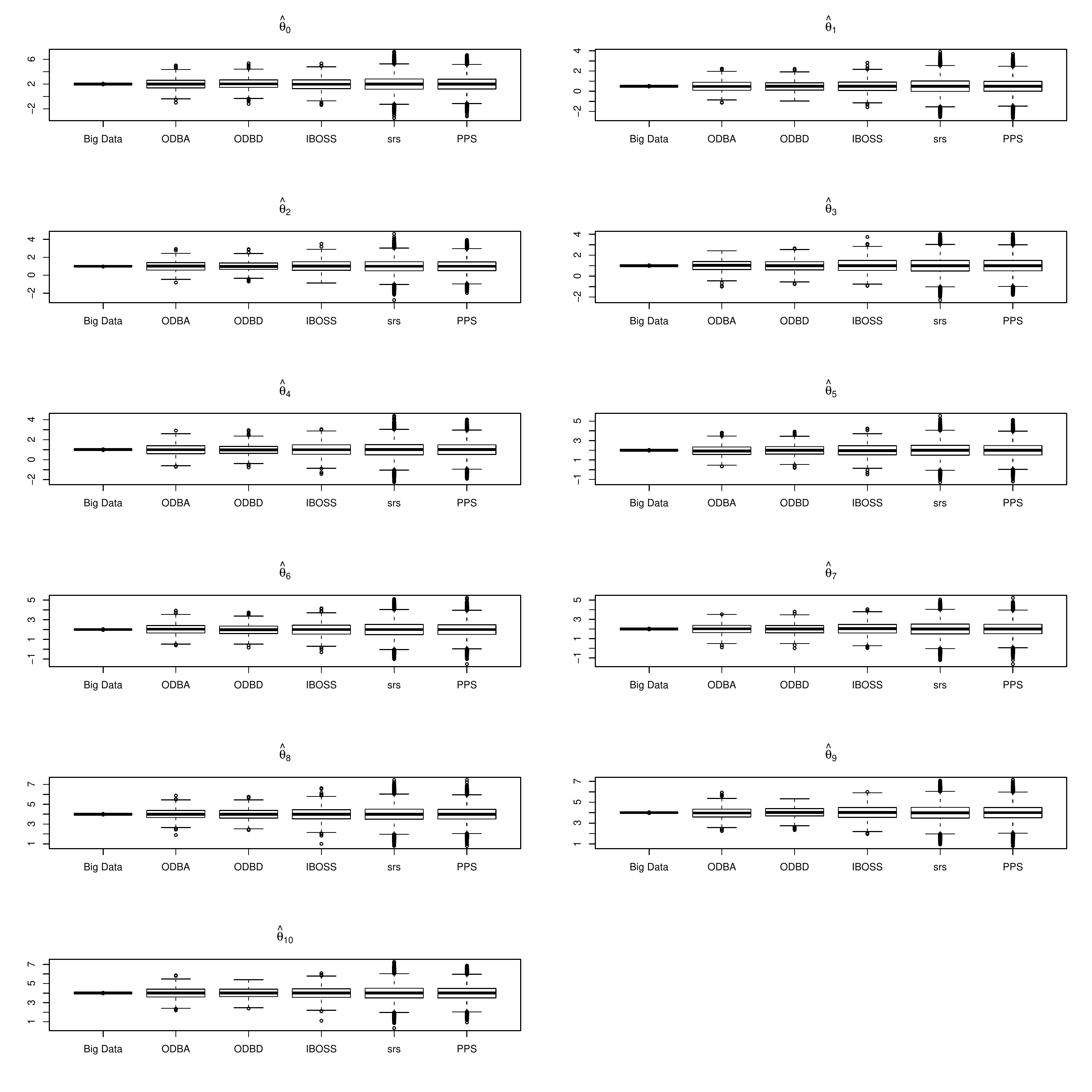}
\end{figure}

The same results concerning the precision of the estimates are displayed in Figure~\ref{fig:3}, which reports the boxplots of the parameter estimates that are obtained in the simulation study. Again, the SSR and PPS methods perform the worst.
\vspace{0.5cm}
{\bf Simulation 2}\\
By adopting the same setting of the previous simulation to generate the design matrix $\mathbf{X}_U$, a second simulation is performed by assuming a logistic model for the response variable. Specifically, a response vector of size $N$ is simulated $R$ times using the following logistic model:
  $$
E(\mathbf{y}_U) = \frac{1}{1+e^{- \mathbf{F}_U \,\theta}}  
$$
with $\mathbf{f}(\mathbf{x})=(1,x_1,x_2, \cdots, x_{10})^T$, $\bm{\theta}=(-1, 1, -0.5, -1, -0.5, 0.25, 2, -0.5, 0.5, 0.5)^T$.
In this case, the D- and A- optimum designs depend on unknown model parameters. We follow the local optimality approach and choose the true value of  $\bm{\theta}$ as the nominal value. At each step $r$, a subsample $s$ of size $n=200$ is drawn from the Big Dataset according to the four different sampling schemes. As before, for each subsample $s$, we compute the D- and A-efficiencies and the ML estimates of the coefficients.

Table~\ref{tab:Eff_log} shows that the IBOSS approach performs worse when the super-population model is non-linear, especially for the A-optimality criterion where it results in the worst efficiency. This is not surprising because IBOSS approach was proposed for linear models; see \cite{Wang:2018b} for the logistic case.

\begin{table*}[!h]
\caption{ Monte Carlo averages of the D- and A-efficiencies of the Big Dataset and the subsamples that are obtained using ODB, IBOSS, SRS and PPS.}
\label{tab:Eff_log}
    \centering
    \begin{tabular}{c|cccccc}
         $\Phi$-Efficiency & Big Data &  ODB & IBOSS & SRS & PPS\\
         \hline
         $D$-Efficiency    & 0.3000 & 0.6234 & 0.3085 & 0.2903 & 0.3616\\
         $A$-Efficiency    & 0.4028 & 0.5468 & 0.0055 & 0.3778 & 0.4475\\
    \end{tabular}
\end{table*}

\begin{table*}[!h]
\caption{Determinants and traces of the Monte Carlo covariance matrix of the estimates of the Big Data set and the subsamples that are obtained using ODB, IBOSS, SRS and PPS.}
    \label{tab:D_A_log}
    \centering
    \begin{tabular}{c|cccccc}
         Criterion & Big Data &  ODB & IBOSS & SRS & PPS\\
         \hline
         Determinant  & 1e-48 &  3e-09 & 1e-05 & 6e-07 & 2e-07\\
         Trace  & 7e-04  & 3.59 & 400 & 4.27 & 4.00\\
    \end{tabular}
\end{table*}

\begin{figure}[!ht]
\vspace{6pc}
\caption{The boxplots of the parameter estimates that are obtained in Simulation 2 using the Big Dataset and the subsamples that are selected using the ODB, IBOSS, SRS and PPS methods.}
\label{fig:4}
\includegraphics[width=0.99\textwidth]{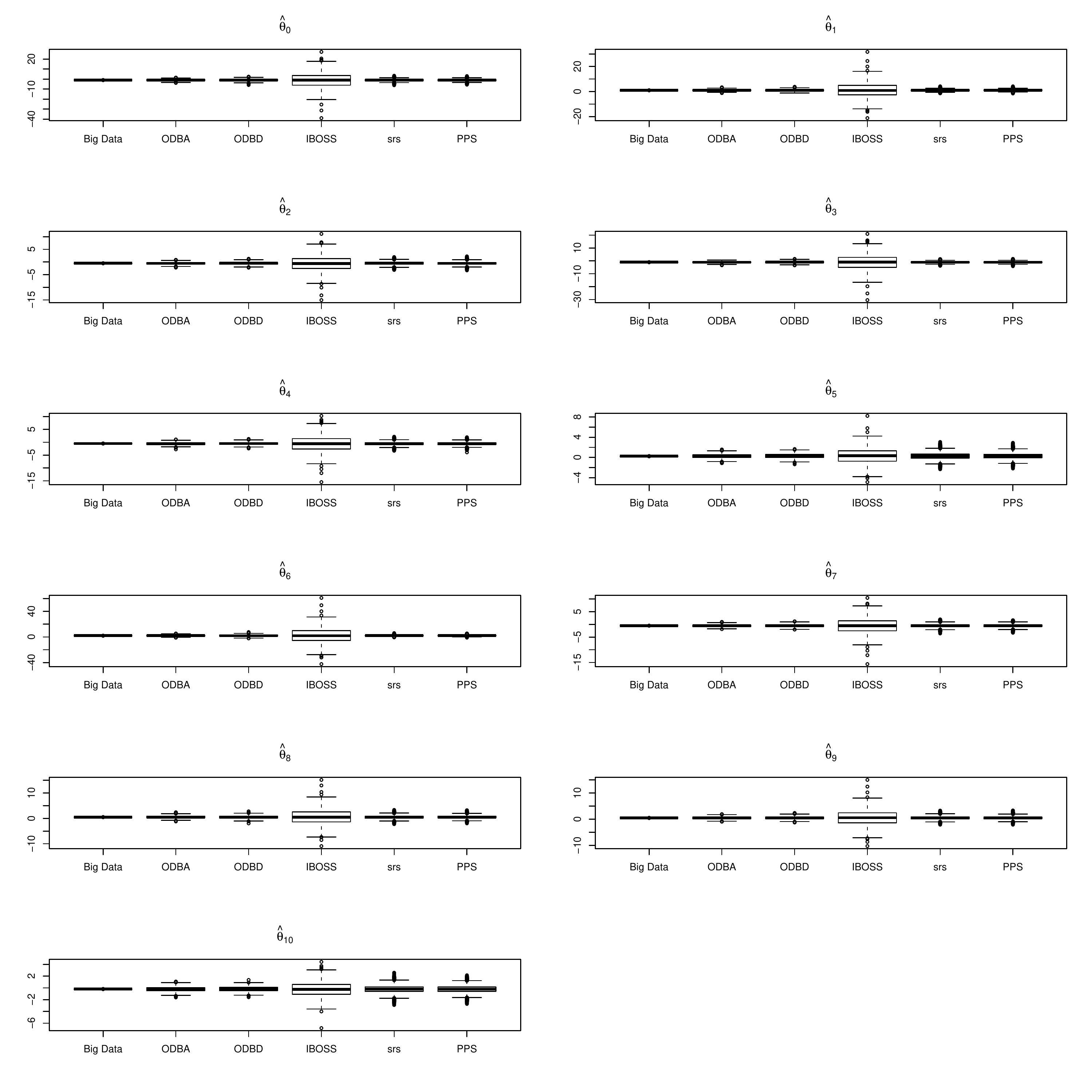}
\end{figure}

Consistently, the largest determinants and traces of the Monte Carlo covariance matrices correspond to the IBOSS approach, as shown in Table~\ref{tab:D_A_log}. Figure~\ref{fig:4}, which displays the parameter estimate boxplots, confirms that the IBOSS algorithm results in the lowest precision.  
The best performance is obtained using the ODB sampling scheme; this result is shown in Figure~\ref{fig:11} in the Supplementary Materials, where IBOSS has been removed to improve the figure quality.

\section{Discussion and conclusions}
In this work, we introduce a new sampling strategy - named ODB - to select a sample from a Big Dataset with $N \gg p$. The method is driven by the optimal design theory of controlled experimentation, which is the standard method of scientific investigation. Actually, our proposal intends to illustrate the advantages of collecting {\em better, fewer} data - instead of {\em big} data - through an optimal experiment. For this reason, we introduce a measure of the quality of the full data set in terms of the per-unit information to illustrate the potential loss in information value when data are not accrued via controlled experimentation.

The simulation study shows that ODB guarantees the selection of the most informative sample to estimate the parameter of the super-population model from which the Big Dataset has been generated.

Furthermore, ODB can be considered to be a quite flexible method. Even if we limited ourselves to D- and A- optimality, ODB could be implemented for other different optimality criteria. Furthermore, through the local linearization approach, ODB can be applied also to GLM models with good results in terms of the efficiency with respect to the competitor selection procedures that are considered in this paper (SRS, PPS, and IBOSS). Unlike IBOSS, ODB is adjusted to guarantee a fixed size $n$ for the sample and it is robust given the permutation order of the covariates. However, this flexibility results in a computation time for Algorithm 1 that is greater than that of IBOSS, which is truly faster. A more optimized code may be achieved.

There are important and unsolved questions that require future study. For the logistic model, the results depend on the nominal value; thus, sensitivity analysis has to be performed to better analyse the performance of our approach. 

Another interesting aspect that we want to explore is the question of sampling from Big Datasets arising from an irregular design space; related works include \cite{Fedorov:1989} and \cite{Pronzato:2006}.

In addition, the impact of a misspecification in  the super-population model - due to selection bias or omitted variable bias - requires further examination.  

We strongly believe that data reduction is an unavoidable aspect of Big data analysis that is required to overcome the  ineffectiveness of some inferential conclusions with large samples. Within this framework, ODB may be a valid approach to extract the most informative data to answer specific questions of interest.

\section*{Acknowledgements}
We are indebted to the participants of the workshop on Model Oriented Design and Analysis (MODA 12) for their useful comments on preliminary versions of this work. Moreover, we are grateful to Luc Pronzato for suggesting some useful references, such as \cite{Wynn:1977b}, \cite{Fedorov:1989} and \cite{Pronzato:2006}.



\newpage
{\bf SUPPLEMENTARY MATERIALS}
\\
\begin{figure}[!ht]

\caption{The full data set of size $N=10.000$ and the subsamples of size $n=120$ when $X_1$ and $X_2$ are independently distributed as $U(-1,1)$, $\mathbf{f}(\mathbf{x})=(1,x_1,x_2, x_1^2, x_2^2, x_1 \cdot x_2)^T$ and A--optimality is used in the exchange algorithm and ODB methods.}
\label{fig:5}
\includegraphics[width=0.99\textwidth]{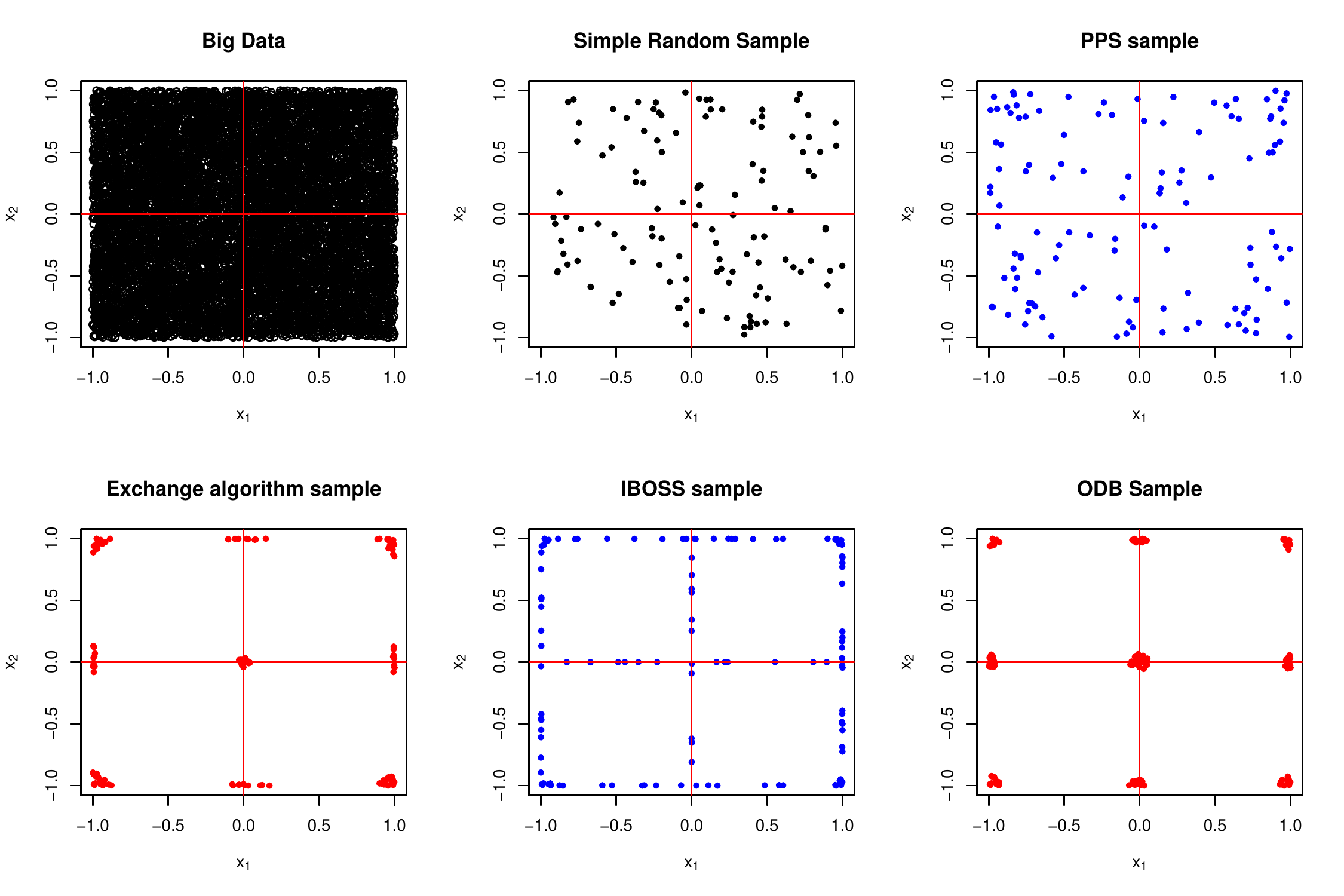}
\end{figure}
\vspace{2pc}

In Figure \ref{fig:5} the A-efficiencies at the different information matrices are:
\begin{eqnarray*}
 && {\rm Eff}_A(\bm{M})=0.441,\quad {\rm Eff}_A(\bm{M}_{SRS})=0.438,\quad {\rm Eff}_A(\bm{M}_{PPS})=0.514,\\
&& {\rm Eff}_A(\bm{M}_{EA})=0.805, \quad {\rm Eff}_A(\bm{M}_{IBOSS})=0.741, \quad {\rm Eff}_A(\bm{M}_{ODB})=0.921.
\end{eqnarray*}

\newpage
\begin{figure}[!ht]
\vspace{6pc}
\caption{The full data set of size $N=10.000$ and the subsamples of size $n=120$ when $(X_1,X_2) \sim N(\mathbf{0},\Sigma)$ 
and D-optimality is used in the exchange algorithm and ODB methods.}
\label{fig:6}
\includegraphics[width=0.99\textwidth]{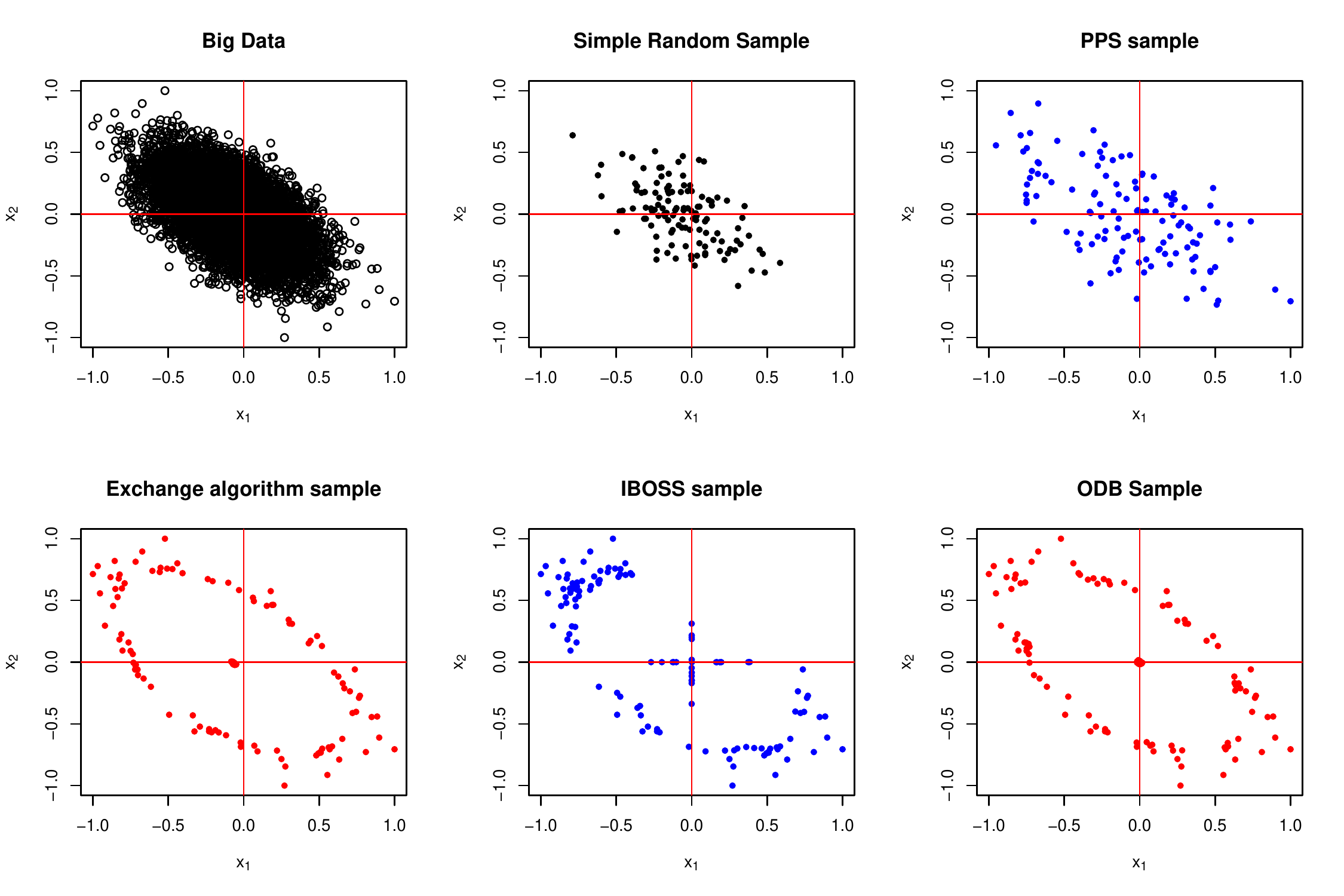}
\end{figure}
\vspace{2pc}

In Figure \ref{fig:6} the D-efficiency  at the different information matrices are: 
\begin{eqnarray*}
&& {\rm Eff}_D(\bm{M})=0.057,\quad {\rm Eff}_D(\bm{M}_{SRS})=0.053,\quad {\rm Eff}_D(\bm{M}_{PPS})=0.161,\\
&& {\rm Eff}_D(\bm{M}_{EA})=0.298, \quad {\rm Eff}_D(\bm{M}_{IBOSS})=0.249, \quad {\rm Eff}_D(\bm{M}_{ODB})=0.282.
\end{eqnarray*}
\newpage
\begin{figure}[!ht]
\vspace{6pc}
\caption{The full data set of size $N=10.000$ and the subsamples of size $n=120$ when $(X_1,X_2) \sim N(\mathbf{0},\Sigma)$ 
and D-optimality is used in the exchange algorithm and ODB methods.}
\label{fig:7}
\includegraphics[width=0.99\textwidth]{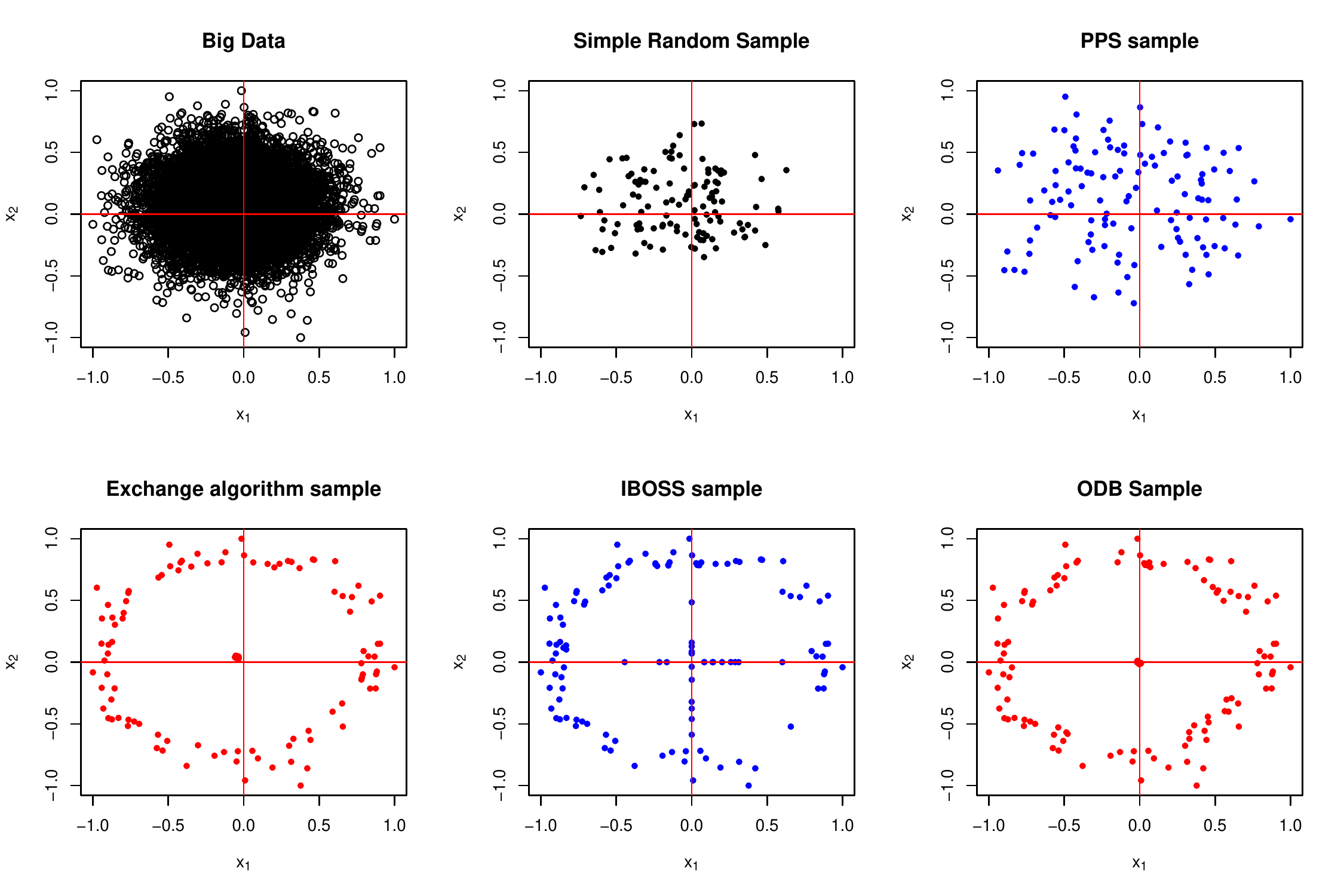}
\end{figure}
\vspace{2pc}

In Figure \ref{fig:7} the D-efficiency  at the different information matrices are: 
\begin{eqnarray*}
&& {\rm Eff}_D(\bm{M})=0.071,\quad {\rm Eff}_D(\bm{M}_{SRS})=0.072,\quad {\rm Eff}_D(\bm{M}_{PPS})=0.193,\\
&& {\rm Eff}_D(\bm{M}_{EA})=0.394, \quad {\rm Eff}_D(\bm{M}_{IBOSS})=0.376, \quad {\rm Eff}_D(\bm{M}_{ODB})=0.378.
\end{eqnarray*}
\newpage
\begin{figure}[!ht]
\vspace{6pc}
\caption{The full data set of size $N=10.000$ and the subsamples of size $n=120$ when $(X_1,X_2) \sim N(\mathbf{0},\Sigma)$ 
and D-optimality is used in the exchange algorithm and ODB methods.}
\label{fig:8}
\includegraphics[width=0.99\textwidth]{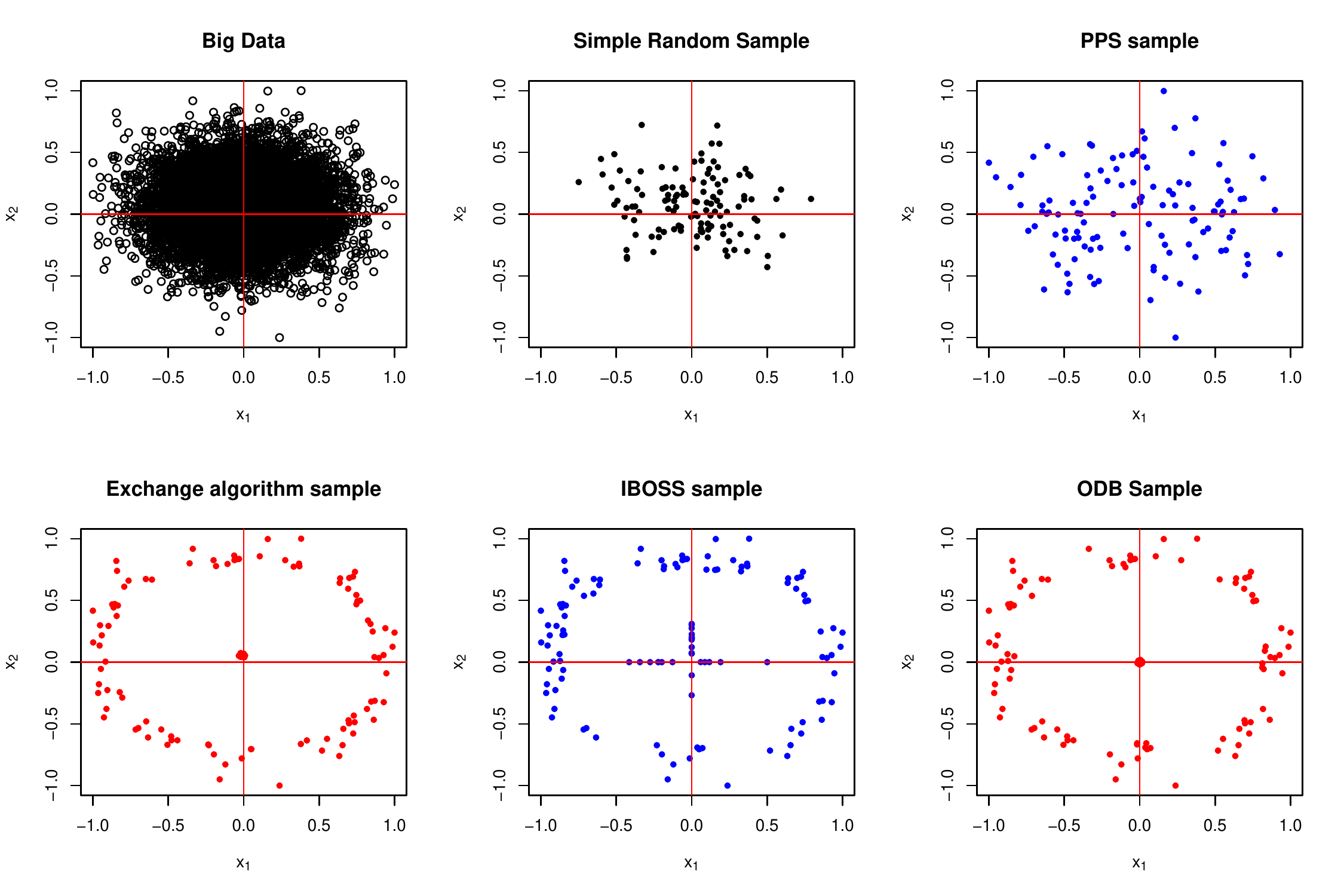}
\end{figure}
\vspace{2pc}

In Figure \ref{fig:8} the A-efficiency  at the different information matrices are: 
\begin{eqnarray*}
&& {\rm Eff}_A(\bm{M})=0.042,\quad {\rm Eff}_A(\bm{M}_{SRS})=0.034,\quad {\rm Eff}_A(\bm{M}_{PPS})=0.146,\\
&& {\rm Eff}_A(\bm{M}_{EA})=0.356, \quad {\rm Eff}_A(\bm{M}_{IBOSS})=0.324, \quad {\rm Eff}_A(\bm{M}_{ODB})=0.355.
\end{eqnarray*}
\newpage
\begin{figure}[!ht]
\vspace{6pc}
\caption{The full data set of size $N=10.000$ and the subsamples of size $n=1000$ when $(X_1,X_2) \sim N(\mathbf{0},\Sigma)$,
 $E(Y)\!\!=\!\!\frac{1}{1+e^{-(\theta_0+\theta_1 x_1+\theta_2 x_2)}}$
 and D-optimality is used in the exchange algorithm and ODB methods. }
\label{fig:9}
\includegraphics[width=0.99\textwidth]{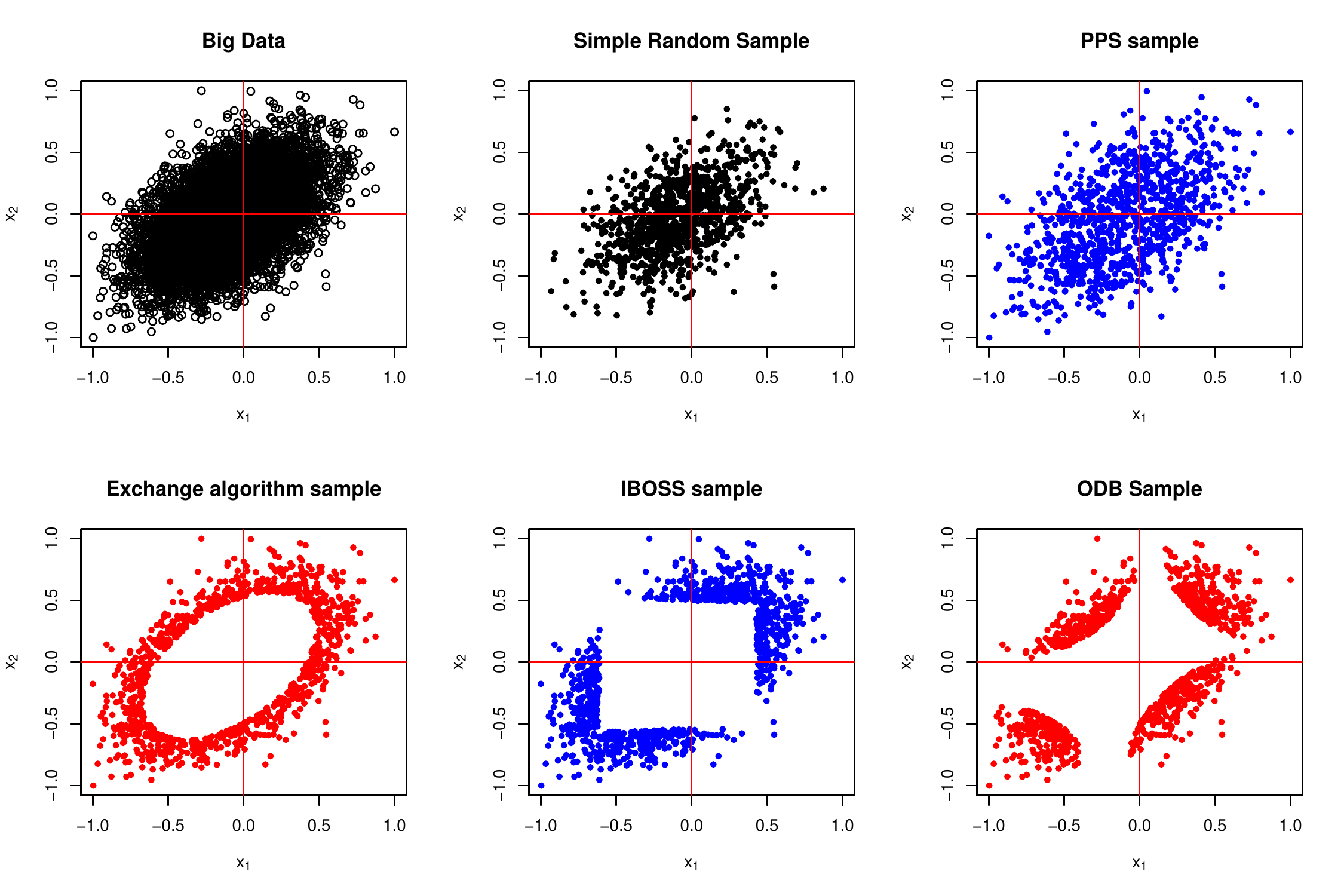}
\end{figure}
\vspace{2pc}
In Figure \ref{fig:9} the D-efficiency  at the different information matrices are: 
\begin{eqnarray*}
&& {\rm Eff}_D(\bm{M})=0.162,\quad {\rm Eff}_D(\bm{M}_{SRS})=0.166,\quad {\rm Eff}_D(\bm{M}_{PPS})=0.230,\\
&& {\rm Eff}_D(\bm{M}_{EA})=0.358, \quad {\rm Eff}_D(\bm{M}_{IBOSS})=0.337, \quad {\rm Eff}_D(\bm{M}_{ODB})=0.336.
\end{eqnarray*}
\newpage
\begin{figure}[!ht]
\vspace{6pc}
\caption{The full data set of size $N=10.000$ and the subsamples of size $n=1000$ when $(X_1,X_2) \sim N(\mathbf{0},\Sigma)$,
$E(Y)\!\!=\!\!\frac{1}{1+e^{-(\theta_0+\theta_1 x_1+\theta_2 x_2)}}$ 
and A-optimality is used in the exchange algorithm and ODB methods.}
\label{fig:10}
\includegraphics[width=0.99\textwidth]{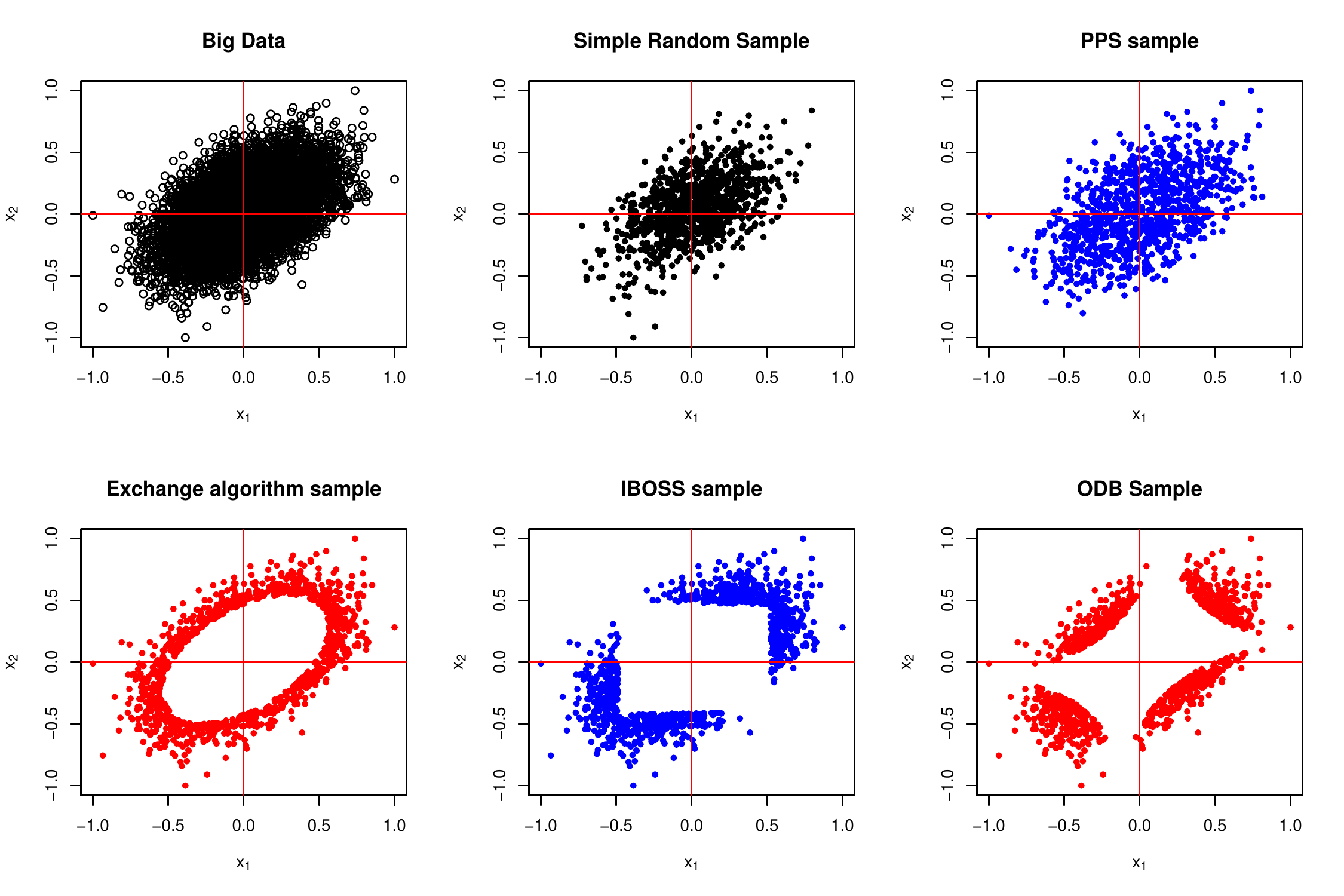}
\end{figure}
\vspace{2pc}
In Figure \ref{fig:10} the A-efficiency  at the different information matrices are: 
\begin{eqnarray*}
&& {\rm Eff}_A(\bm{M})=0.069,\quad {\rm Eff}_A(\bm{M}_{SRS})=0.071,\quad {\rm Eff}_A(\bm{M}_{PPS})=0.105,\\
&& {\rm Eff}_A(\bm{M}_{EA})=0.216, \quad {\rm Eff}_A(\bm{M}_{IBOSS})=0.169, \quad {\rm Eff}_A(\bm{M}_{ODB})=0.198.
\end{eqnarray*}
\newpage

\begin{figure}[!ht]
\vspace{6pc}
\caption{The boxplots of the parameter estimates that are obtained in Simulation 2 using the Big Dataset and the subsamples that are selected using the ODB, SRS and PPS methods.}
\label{fig:11}
\includegraphics[width=0.99\textwidth]{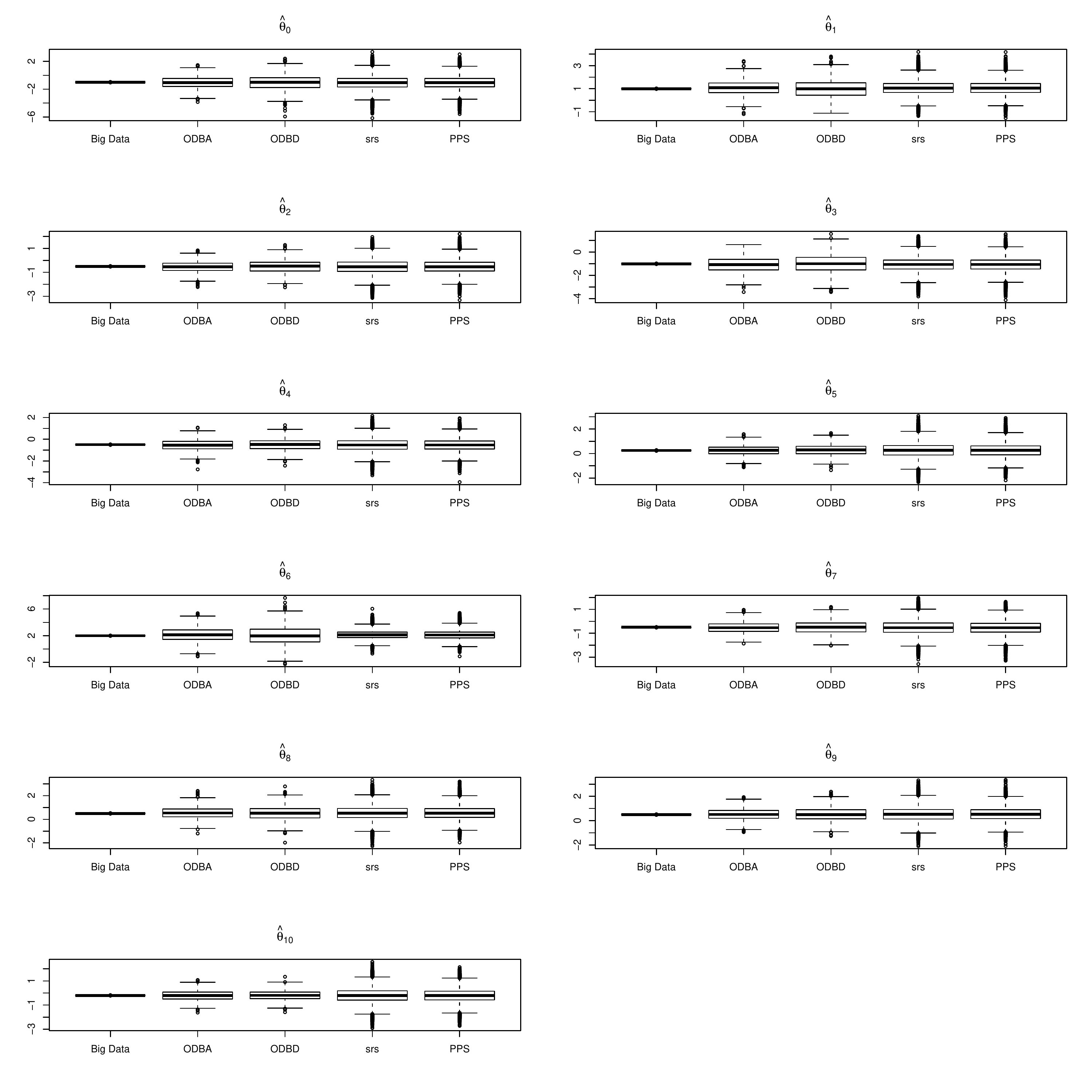}
\end{figure}

\end{document}